\documentclass[letterpaper,twocolumn,10pt]{article}

\usepackage[latin1]{inputenc}
\usepackage[english]{babel}
\usepackage{scs}
\usepackage{graphicx}
\usepackage{amssymb,amsmath}

\usepackage{algorithm}
\usepackage{algorithmic}

\begin{document}

\title{Strategic Evolution of Adversaries Against Temporal Platform Diversity Active Cyber Defenses}
\author{
Michael L.~Winterrose and Kevin M.~Carter \\
MIT Lincoln Laboratory, 244 Wood Street, Lexington, MA 02421, USA \\
\{michael.winterrose, kevin.carter\}@ll.mit.edu
}

\maketitle


\begin{abstract}
\noindent
Adversarial dynamics are a critical facet within the cyber security domain, in which there exists a co-evolution between attackers and defenders in any given threat scenario. While defenders leverage capabilities to minimize the potential impact of an attack, the adversary is simultaneously developing countermeasures to the observed defenses. In this study, we develop a set of tools to model the adaptive strategy formulation of an intelligent actor against an active cyber defensive system. We encode strategies as binary chromosomes representing finite state machines that evolve according to Holland's genetic algorithm. We study the strategic considerations including overall actor reward balanced against the complexity of the determined strategies. We present a series of simulation results demonstrating the ability to automatically search a large strategy space for optimal resultant fitness against a variety of counter-strategies. \footnote{This work is sponsored by the Department of Defense under Air Force Contract FA8721-05-C-0002. Opinions, interpretations, conclusions and recommendations are those of the author and are not necessarily endorsed by the United States Government.}
\end{abstract}

\section{Introduction}

Adversarial dynamics are at the heart of many important security challenges. The central issue in this problem class is the existence of an intelligent, adaptable adversary able to actively counter defensive moves. Models developed to increase understanding in these areas must include adaptive learning as a central component if they aim to capture the adversarial dynamics observed to dominate many real-world conflict scenarios \cite{cybenko-2013}.  

Cyber security researchers, attempting to understand and predictably influence the highly dynamic, noisy, and innovative cyber realm, require tools that shed light on strategic interactions that are often out of equilibrium and driven by adaptive learning. Furthermore, the sought after techniques should be able to account for the limited ability of real world actors to gather and process information and the consequent bounded nature of the rationality that characterizes real world adversarial opponents. 

In recent years \emph{active cyber defense} techniques have been an important area of research and development \cite{colbaugh-2012a, okhravi-2013}. A prominent example of a family of active defense techniques that have recently been under development is that of moving target defenses \cite{jajodia-2011, jajodia-2013}, which systematically vary a system's attack surface to diminish the inherent advantages cyber attackers typically enjoy \cite{jajodia-2013}. Temporal platform migration systems \cite{talent} are a class of moving target defenses that dynamically change the system interface (e.g.~operating system (OS), spam filter, etc.) over time. These techniques work under the assumption that the attacker has limited resources and generally does not have exploits for \emph{all} platforms at his disposal. As such, migrating between platforms with some frequency reduces the ability for an attacker to maintain persistence on a system. Additionally, it increases the uncertainty for an attacker that aims to expend resources towards countermeasure development.

Recent studies have examined the optimal scheduling policy for a temporal platform migration moving target defense. According to \cite{Colbaugh-2012}, a uniform random scheduling of active spam filters gave superior performance against an evolving adversary. On the other hand, \cite{Carter-2013} found that a scheduling policy that maximizes the diversity of the platforms played in each successive round as quantified by a measure of platform similarity was superior. Crucially, the threat model in these studies differed, with \cite{Carter-2013} modeling a threat that required attacker persistence in the system before a payoff accrued, in contrast to \cite{Colbaugh-2012} which used a simple attacker model with no requirement for persistence.

The primary aim of this study is not to formulate new moving target defense strategies, but rather to develop a set of modeling tools that facilitate insight into the effect these strategies have on the potential success and strategy evolution of a cyber attacker that is able to observe a system and react adaptively to defensive moves. In this study we demonstrate a tractable approach to the modeling of adaptive strategic learning by boundedly rational agents and apply the scheme for the first time to a resource allocation problem in the cyber security domain.   

The modeling approach we employ involves the use of genetic algorithms, which have been used in a number of security applications and in studies of agent strategy formulation. Intrusion detection systems have incorporated genetic algorithms \cite{crosbie-1995, dasgupta-2001, pillai-2004, li-2004}. Genetic algorithms have also been used to create active defense strategies \cite{caltagirone-2005} and to generate diverse configurations in a moving target system \cite{crouse-2011, crouse-2012}. In his pioneering work Axelrod \cite{axelrod-1987} used a genetic algorithm to explore strategy evolution in the Iterated Prisoner's Dilemma (IPD). Miller \cite{miller-1996} (see below) brought together the ideas of Axelrod with methods from automata theory \cite{aumann-1981, rubinstein-1986, rubinstein-1988} developed to model the strategic play of boundedly rational agents. The combination of a genetic algorithm and finite state machines allows for the tractable modeling of boundedly rational agents able to learn and adapt in dynamic environments \cite{miller-1996}. 

 The major contributions of this work are as follows:
\begin{enumerate}
\item We present the first use of genetic algorithms to study optimal attacker strategies against temporal platform diversity defenses
\item We quantify a \emph{strategic complexity} measure to balance the cost/benefit within resource investment
\item We study the efficacy of different defender strategies against an evolving adversary
\end{enumerate}

The paper proceeds as follows: In Section \ref{S:Methods} we formulate the problem and present our techniques and evaluation metrics. In Section \ref{S:Experiments} we present a series of experiments in which the adaptive attacker interacts with different defender strategies, finally concluding by discussing future work in Section \ref{S:Conclusion}

\section{Problem and Methods}\label{S:Methods}

\subsection{Attacker/Defender Game Scenario}

We study the interaction between the attacker and defender through the dynamics of a two-player game. In our scenario the attacker observes the operating system the defender, utilizing a moving target defense, activates in each round of play. The attacker uses this information to formulate a strategy for resource investment, with the goal to bring exploits (e.g.~countermeasures) into existence to most effectively compromise the defender's system and maximize his own gain.

In each match (i.e.~round of play), the attacker uses any exploits it has developed against the activated platform. Success for the attacker in a match occurs when he has an available exploit that works against the platform activated by the defender. Intuitively, the attacker gains a reward if they are able to compromise the defender's system during the match, and earns nothing otherwise. One may view this as the ability for the attacker to exfiltrate data, perform reconnaissance, steal credentials, etc.; all attacks that benefit from system access. 

Exploits are developed by cumulative attacker investment; in each round of play the attacker invests a single resource to the creation of a chosen exploit. The amount of investment required to bring a given exploit into existence is randomly determined at the beginning of each game. The attacker is not informed \emph{a priori} of the number of resources that will be required to bring a given exploit into existence, instead discovering this fact only after having successfully created the exploit through the required allocation of resources. This uncertainty matches reality in that a real-world attacker is not in general able to predict \emph{a priori} the level of effort that will be required to develop new exploits against a given system. Without loss of generality, in this initial version of the game, there exists one possible exploit for each type of platform a defender might utilize. 

All exploits in this study are assumed to be \emph{zero-day} exploits, meaning that they are unobserved by the defender when used against the corresponding operating system. Intuitively, the defender has no detection capability against an attack they have no knowledge of. This has ramifications in the sense that the defender has no means or reason to change their own defensive strategy, and hence will stick to the pre-assigned strategy. 

We note that while the agents in agent-based models typically manifest attributes such as heterogeneity, adaptation, interaction, autonomy, and situatedness \cite{tolk-2009, yilmaz-2009}, our current game scenario focuses primarily on one of these attributes, namely adaptive strategy development. Our intention in this study is to focus on developing a powerful set of tools for modeling adaptive strategy evolution in cyber agent systems. These tools will be incorporated into more complete agent contexts in future studies.

\subsection{Moore Machine Strategy Encoding}

A finite-state automata or finite-state machine is an abstract machine that takes discrete inputs from an environment and specifies a discrete output in response. An agent modeled by a finite state machine will occupy only one state at any point in time. Such an agent transitions between states based on triggering events. A Moore machine \cite{moore-1956} is one kind of finite-state automata in which the output from the machine depends solely on the machine's internal state. 

A Moore machine is formally defined by four components, $<Q, q_{0}, \Lambda, \xi >$ , where Q is the set of possible internal states,  $q_{0} \in Q$ is the internal state the machine begins in, $\Lambda$ is a function mapping the internal states to actions in the game being played, and $\xi$ is the transition function mapping an internal state of the machine and the observed move of the opponent into the next internal state that will become active. Moore machines provide a simple, yet rich and flexible tool with which to encode and evolve game strategies \cite{miller-1996}.

\begin{figure}[t]
\center
\includegraphics[width=6.0cm]{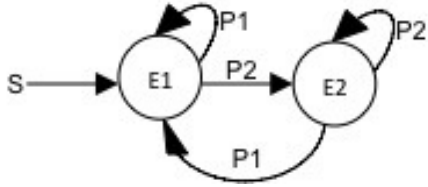} 
\caption{Transition diagram for a possible attacker strategy. In this example the attacker invests in the creation of platform zero-day vulnerabilities with a frequency proportional to the platforms it observes the defender to activate in the rounds of play.}
\label{fsmPic}
\end{figure}

Figure \ref{fsmPic} shows the transition diagram for a possible strategy for the game studied in this work represented as a Moore machine. Here $s$ marks the state the machine begins a game in. States are labeled by $\Lambda : Q \rightarrow A_{i}\in\{E_{1}, E_{2}, . . . E_{n} \}$, where $E_{n}$ indicates the attacker devotes his zero-day exploit creation resource in the current round to the creation of an exploit for platform $n$. The transitions in Fig.~\ref{fsmPic} are labeled by $\{P_{1}, P_{2}, . . .P_{n} \}$, indicating that the attacker transitions to its next active state based on the platform $P_{n}$ it observes the defender to have made active in the current round.

To summarize, the attacker makes an initial investment $E_{n}$ in the creation of zero-day exploit for platform $n$, observes the platform type the defender has activated in the current round of its moving target defense, and selects the next investment in exploit creation to make based on this information.

\subsection{Genetic Algorithm}

The central issue this simulation model aims to address is the efficiency and effectiveness with which an adversary learns to counter various moving target defense strategies. Learning and adaptation is modeled here using a genetic algorithm \cite{holland-1975}. The genetic algorithm is a robust search method that does not require its objective function to be linear or continuous and has proven itself efficient at searching large fitness landscapes. The algorithm used in this study is outlined in Algorithm \ref{algo1}.

 \begin{algorithm}[t]
\caption{Implemented Genetic Algorithm}
\label{algo1}
\begin{algorithmic}
\STATE 1) At generation $g$=1, randomly initialize $N$ strategies, \\
\STATE  \ \ \  $i$=1 to $N$.
\STATE 2) Generate a fitness score $F_{i,g}$ for each strategy based on \\ 
\STATE \ \ \  its success against the defender.
\STATE 3) Form a new population of $N$ structures.
\STATE \ \ \ a) Create $0.6N$ new strategies. 
\STATE \ \ \ \ \ \ \ i) Select 2 parents using the tournament selection 
\STATE \ \ \ \ \ \ \ \ \ \ algorithm (with replacement).
\STATE \ \ \ \ \ \ ii) Form 2 children from the two selected parents \ \ 
\STATE \ \ \ \ \ \  \ \ \ \  through crossover.
\STATE \ \ \ \ \ \ \ iii) Repeat (i)  (ii) until $0.6N$ new structures are \\
\STATE \ \ \ \ \ \ \ \ \ \ formed.
\STATE \ \ \ b) Select $0.4N$ strategies from the old population using \\
\STATE \ \ \ \ \  tournament selection and copy them into the \\
\STATE \ \ \ \ \  new generation.
\STATE 4) Mutate (see below) the new generation of strategies
\STATE 5) Increment $g$ by 1 (next generation), and iterate (go \\ 
\STATE \ \ \ \ to Step 2). 
\end{algorithmic}
\end{algorithm}

\vspace{ 3 mm }

In this initial version of the study the defender chooses one operating system to make active in each round from its supply of two operating systems. We call this initial version of the attacker-defender contest the \emph{binary game}. We denote the operating systems available to the defender as \emph{OS-A} and \emph{OS-B}. As stated previously the attacker  can create a zero-day exploit effective against each operating system available to the defender. We denote these \emph{ZD-A} and \emph{ZD-B}. 

\begin{figure}[t]
\center
\includegraphics[width=9.0cm]{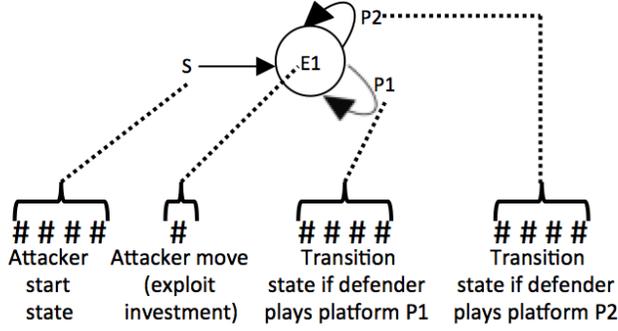} 
\caption{Chromosomal structure for an attacker strategy consisting of a single state. Chromosome bits are represented by $ \# \in \{0, 1\}$. Each attacker state is represented in the chromosome by 9 bits, 1 bit for the action specification, and 4 bits each for the two transition states in this example. Binary encoding is used throughout.}
\label{automChromoMap}
\end{figure}

The Moore machine is mapped into a binary chromosome according to the scheme described in  \cite{miller-1996}. We limit the Moore machine representation to 16 states,  following \cite{miller-1996}. Auxiliary studies show limiting the Moore machines used in the study to 16 states does not strongly affect strategy development in our simple game scenario. Each state of the 16 state Moore machine maps to 9 bits of a binary chromosome. One of the bits specifies the exploit type that will be invested in when the attacker occupies the state. The subsequent 8 bits specifies the two transitions (4 bits for each transition encoded in a binary representation) the attacker can make out of the state based on its observation of the operating system the defender makes active in the current round. This is depicted in Fig. \ref{automChromoMap}. An additional 4 bits are added to the chromosome to specify the starting state for the machine. These factors together lead to the $148$ bit chromosome used in this study to encode the attacker strategy.

This encoding scheme has been previously used to study strategy co-evolution in the Iterated Prisoner's Dilemma (IPD) \cite{miller-1996, ho-1996}. To the best of our knowledge, this study marks the first application of this encoding scheme developed in the IPD literature to a resource allocation problem, and its first application to a cyber security problem.

A \emph{single-point} crossover operation is used in this study. In this operation, a single crossover point $\zeta \in \{1,2, . . . 148\}$ on each of two parent chromosomes is selected at random. The first child combines the first $\zeta$ bits from the first parent with all bits after the $\zeta+1$ chromosome position of the second parent to form a new chromosome. The second child takes all bits after the $\zeta+1$ chromosome position from the first parent and combines it with the first $\zeta$ bits of parent two's chromosome to form a new strategy. 

The mutation operation is implemented here such that each chromosomal position on each of the strategy population members has a low probability of mutation in each generation, with the probability being specified by a \emph{mutation rate} parameter in the model. The mutation rate is selected in our study such that on average half of the chromosomes experience a single bit flip in each generation.

\subsection{Model for Attacker Fitness}

Attacker fitness  determination takes into account three aspects of the attacker's play against the defender. These are game play success ($G$), exploit creation success ($C$), and strategic complexity costs ($S$). In generation $g$ attacker $i$ has a fitness function of the form,

\begin{equation}
F_{i,g} = G_{i,g} + C_{i,g} - S_{i,g}.
\label{fitnessEqt}
\end{equation}

Game payoff for attacker $i$ in generation $g$ is expressed in terms of the attacker's success ($\Phi_{i,j}$) in compromising the system with its created zero-day exploits in the course of play against defender $j$,

\begin{equation}
G_{i,g} = \sum_{j=1}^{T_{g}} \Phi_{i,j}, 
\label{payoffEqt}
\end{equation} 

\noindent
with $\Phi_{i,j}$ set to $1$ if the attacker successfully compromises defender $j$'s system in a given match, and $0$ otherwise. $T_{g}$ is the total number of matches in a given generation between attacker $i$ and defender $j$. For example, if in a $365$-match game the attacker compromises the defender's system for $50$ matches, then the $G_{i,g}$ term is set to $50$ in Eqt. \ref{fitnessEqt}.

The $C_{i,g}$ term in Eqt. \ref{fitnessEqt} models the notion that each exploit created brings an intrinsic reward, $\delta$, to the attacker, independent of their subsequent use in actual attacks. After examining the effect a range of possible values for $\delta$ have on attacker strategy, we set $\delta$ equal to a value of $1$ in this study. This value encourages balanced investment by the attacker while directly impacting overall fitness scores minimally. If $Z_{g}$ exploits are created by attacker $i$ in generation $g$ we have,

\begin{equation}
C_{i,g} = \sum_{k=1}^{Z_{g}} \delta_{i,k}.
\label{creationEqt}
\end{equation}

\noindent
As an example, if $2$ exploits are created in the course of an attacker-defender game, we have $C_{i,g}\,=\,2\delta_{i}=2$, using our chosen $\delta$ value. 

The third fitness criteria is the strategic complexity ($S_{i,g}$) of attacker $i$'s exploit investment policy. We quantify the complexity cost as,

\begin{equation}
S_{i,g} = \beta \, \gamma \, \tau_{i,g}.
\label{stratcomplEqt}
\end{equation}  

\noindent
Here $\tau_{i,g}$ is the number of transitions attacker $i$ makes between Moore machine states in the course of its matches against the defender in a given generation $g$. This is multiplied by $\gamma$, the \emph{transition-penalty} term, which we set to $max[\Phi_{i,j}]$, and the \emph{unit strategic complexity} $\beta$, ranging between $0$ and $1$. The value of $\beta$ can be adjusted to emphasize or de-emphasize the importance of strategic complexity in the fitness evaluation. The strategic complexity term in Eqt.~\ref{fitnessEqt} is of a similar form to that used in \cite{ho-1996} for characterizing complexity costs in the IPD. 

A less complex strategy is one with a simpler transition structure \cite{banks-1990}, i.e.~given a certain level of success against the defender, a strategy that causes the Moore machine encoding the attacker strategy to transition less often between states is preferable. We leverage this to imitate the natural cost/benefit analysis that any human actor factors into strategy determination; a slightly less rewarding strategy is often preferable if it is significantly less complex. As a baseline value, we find that setting $\beta\,=\,0.1$ biases evolved strategies toward simplicity without causing strong distortions in game play policy. We adopt this value throughout the study. Taking this value for $\beta$ we find that an attacker that makes $30$ transitions between Moore machine states in the course of a $365$-match game pays a strategic complexity cost of 3 points, for example. 

\subsection{Investment Bias Metric} \label{InvBiasSection}
 
We define the \emph{Investment Bias}, $\Gamma$, the normalized difference between the mean investment in the creation of ZD-A versus ZD-B exploits, as a measure of the inequality in attacker resource investment,
\begin{equation}
\Gamma = \frac  {\langle I_{ZDB} \rangle  - \langle I_{ZDA}  \rangle}  {\langle I_{ZDB} \rangle  + \langle I_{ZDA}  \rangle},
\label{OrderParamInvestEqt}
\end{equation}
\noindent
where $<I_{ZD}>$ is the average investment in a zero-day exploit ($ZD$) over a population of $N$ attacker strategies and a number of simulation runs carried out to account for stochasticity in the model. Particular emphasis is placed on the investment bias in the analysis below because attacker exploit creation investment patterns directly reflect the learned attacker strategy. Alternative measures of strategy are dependent on factors exogenous to learned attacker strategy such as exploit creation cost values. Attacker investment bias offers the most direct insight into attacker strategy.

\section{Experiments}\label{S:Experiments}

\subsection{Simulation Set-Up}

For ease of interpretation, and without loss of generality, we limit the defender to 2 platforms in this study. Each parameterization of the simulation was allowed to evolve for 100 generations, with a single generation consisting of each of the $N=30$ attacker strategies in the population playing $T=365$ matches against the defender. The 100 generation run was repeated 100 times to account for stochasticity in the model. Simulations were created and run in the NetLogo modeling environment \cite{wilensky-1999}.

\begin{figure}[t]
\center   
\includegraphics[width=9.0cm]{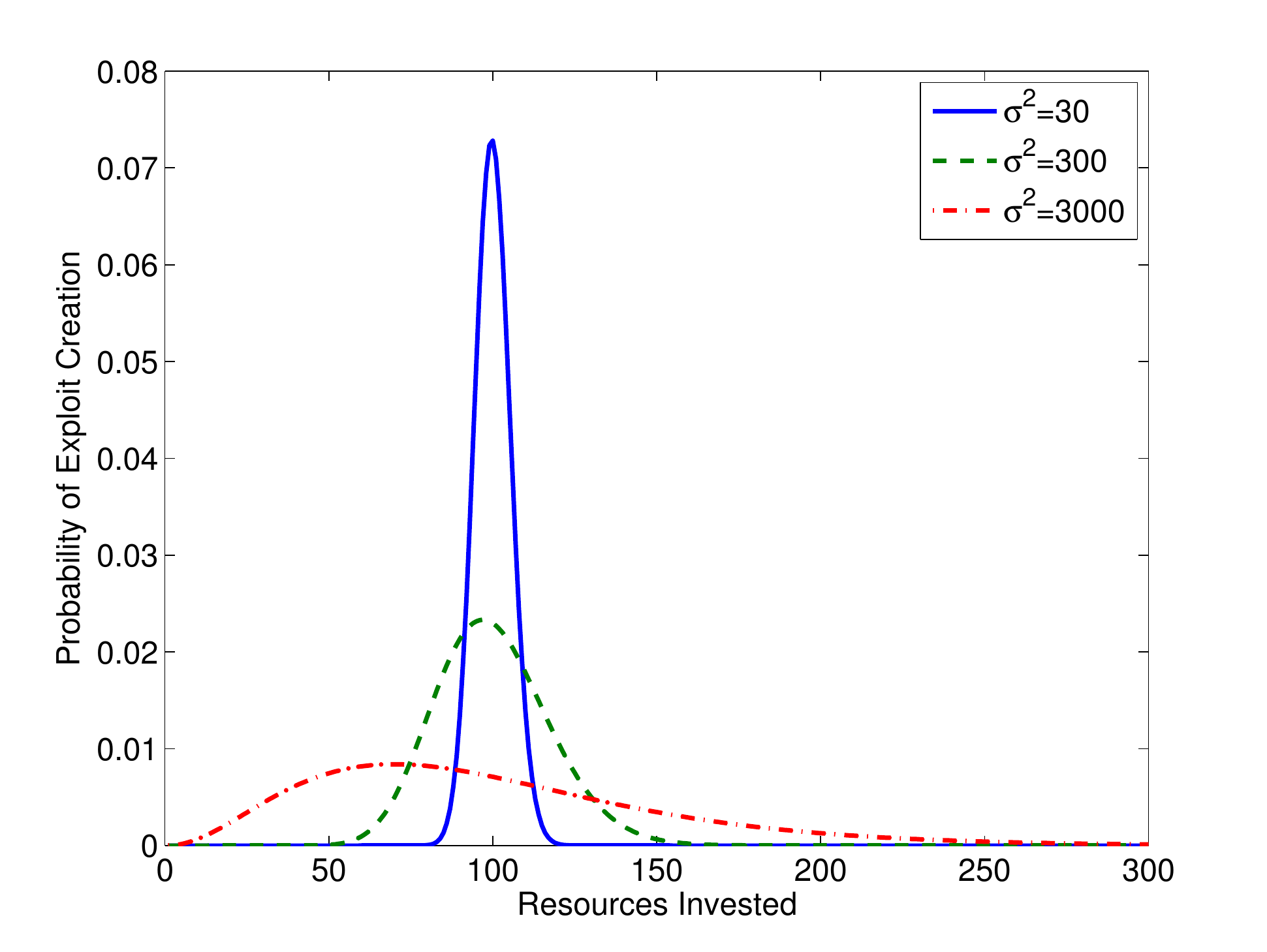}   
\caption{Example Gamma distribution parameterizations with $\mu=100$ and $\sigma^{2}$ set as indicated.}
\label{gammaPic}
\end{figure}

Recall that the attacker resource investment required to bring a given zero-day exploit into existence is stochastic and unknown to the attacker. We determine this cost by sampling a Gamma distribution parameterized by shape parameter $\alpha$ and rate parameter $\lambda$. The shape and rate parameters can be written in terms of a mean ($\mu$) and variance ($\sigma^{2}$) according to the relations,
\begin{equation}
\alpha = \frac{\mu^2}{\sigma^2}, \lambda = \frac{\mu}{\sigma^2}.
\label{alphaEqt}
\end{equation}
\noindent
The parameterization of the Gamma distribution for zero-day costs is set at the beginning of each generation of genetic algorithm evolution through the choice of a mean and variance value. The effect parameterization choices has on the resulting distribution of exploit creation costs is shown in Fig.~\ref{gammaPic}. For the duration of this study, we set $\mu=100$ and $\sigma^{2}=30$ for exploit creation.

\subsection{The 1-to-1 Game}

We denote by the \emph{1-to-1 Game} a series of contests in which the attacker faces defenders that activate OS-A and OS-B with approximately equal frequency in each contest. We use the term \emph{contest} and the term \emph{game} interchangeably to denote a set of $T$ matches that a single attacker plays consecutively against the defender. While within a game the frequency of OS-A and OS-B activation is equal, the distribution of these activations among the $T$ matches differs among defender types within the \emph{1-to-1} defender family. 

The \emph{SingleFlip-FixedOrder} defender activates OS-A in 182 consecutive matches, then switches to OS-B for the next 183 matches. The defender denoted by \emph{SingleFlip-RandomOrder}  plays a similar strategy, but instead of predictably playing OS-A for half the matches then switching to OS-B, this defender instead decides uniformly at random at the start of each game whether to begin play with OS-A or OS-B. The \emph{EachMatchFlip-FixedAlternating} defender plays OS-A followed by OS-B, followed by OS-A, then OS-B again, and so on, alternating between activating each operating system on each consecutive match. Finally, the \emph{EachMatchFlip-RandomOrder} defender chooses at random on each match whether to activate OS-A or OS-B.

\begin{figure}[t]
\center
\includegraphics[width=9.0cm]{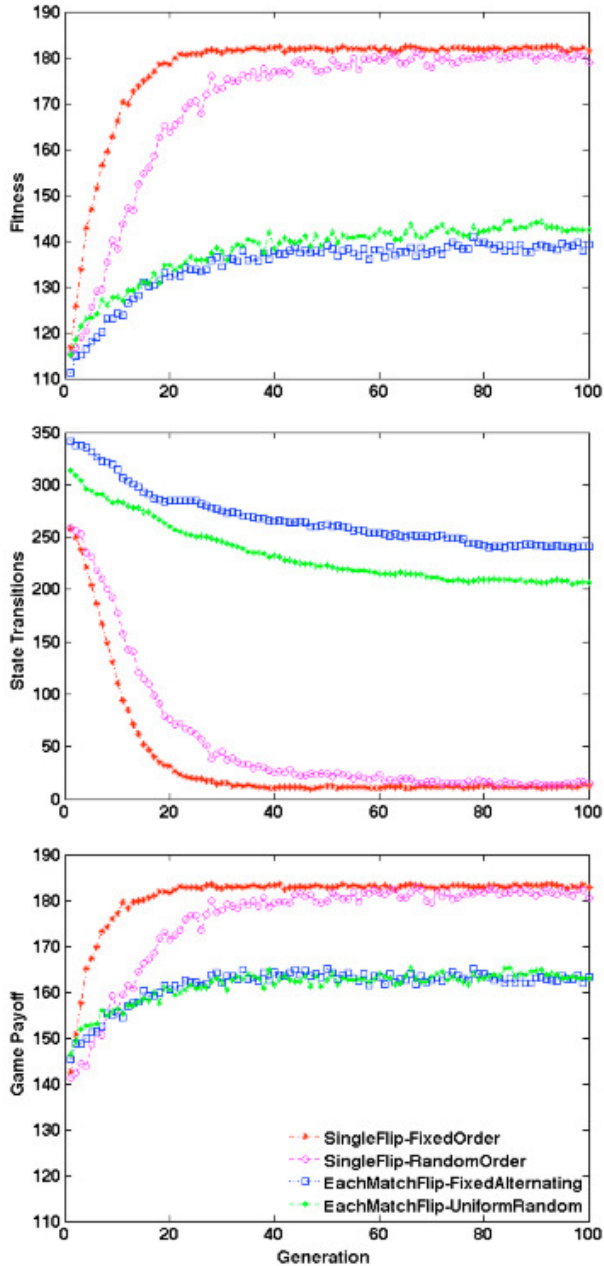} 
\caption{Attacker fitness (upper), state transitions (middle), and game payoff (lower) averaged over the 30 agent population and 100 simulation runs for the 4 defenders playing \emph{1-to-1} strategies.}
\label{1to1Triple}
\end{figure}

\begin{figure}[t]
\center
\includegraphics[width=9.0cm]{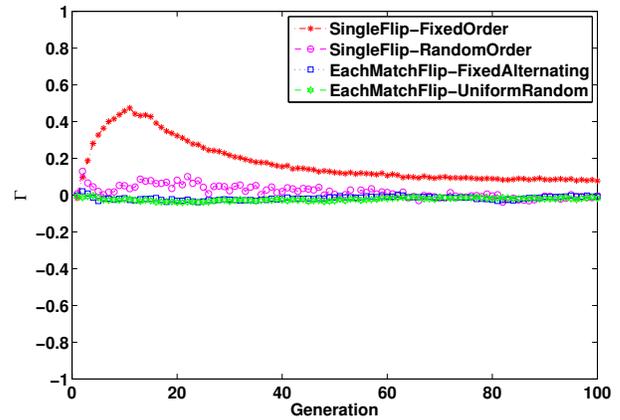} 
\caption{Attacker investment bias ($\Gamma$) for each of the platform migration policies in the \emph{1-to-1} defender family.}
\label{invBias-1to1}
\end{figure}

Figure \ref{1to1Triple}  shows the mean attacker fitness as well as the strategic complexity cost and game payoff results as a function of the genetic algorithm generation. The attacker is able to evolve more effectively against the \emph{SingleFlip} defenders in which the defender predictably activates one of the operating systems available to it during each half of match play. Figure \ref{invBias-1to1} shows the investment bias ($\Gamma$) (see Section \ref{InvBiasSection}) for these \emph{1-to-1} cases. The frequency of defender OS activations within a game strongly influences the investment bias of the attacker's learned strategy, with the \emph{1-to-1} defender strategies causing the attacker to evolve strategies with investment bias clustered around zero, indicating balanced investment in ZD-A and ZD-B creation over the course of a game. The deviation from this pattern in $\Gamma$ is for the \emph{SingleFlip-FixedOrder} defender, which causes the attacker to invest more heavily in ZD-A. 

A clearer picture of learned attacker strategy can be gained from understanding the \emph{match-level} choices the attacker is making. We find that when the defender predictably plays a monolithic OS-B strategy in later rounds of the game the attacker quickly (generationally) learns to predict and exploit this regularity, ignoring the defender's early activations of OS-A and focusing investment on the creation of ZD-B for use in the second half of the game. Note that this learned strategy by the attacker is also taking account for the approximate cost of zero-day exploit creation in that the attacker has learned after a few generations that it needs to begin investing in  the creation of ZD-B well before the defender begins activating OS-B in order to gain full benefit from the defender's predictable behavior. Also note that while the attacker does learn the approximate cost of exploit creation (the mean cost of exploit creation is 100 units in this example), the attacker seems to believe the cost is approximately 182 units, only switching to invest in the creation of ZD-A after the defender transitions from playing OS-A to OS-B at the 182nd round. This inability of the attacker to learn to predict the true exploit creation costs was found to be a robust feature of the learning algorithms used for the present game. We believe this observation is related to the fact that the attacker uses all available exploits against the defender's system on each match, which we speculate makes the attacker insensitive to the precise effect the play of each individual exploit is having on the game payoff. This is a topic we plan to revisit in future studies which will look at other variants of evolutionary algorithms and game structure to understand more fully why the current attacker consistently fails to distinguish the true costs of exploit creation from the ratio of defender play of said exploits.

A portion of the superior performance of the attacker against the \emph{SingleFlip} defenders can be traced to superior game-play strategies for these \emph{SingleFlip} cases (Fig. \ref{1to1Triple} (lower)), and a portion can be traced to the ability of the attacker to discover strategies of minimal complexity costs (Fig. \ref{1to1Triple} (middle)). Future studies will examine the robustness under $\beta$ variation of strategic complexity cost differences for attackers facing \emph{SingleFlip} versus \emph{EachMatchFlip} defenders. In the case shown in Fig. \ref{1to1Triple} it makes intuitive sense that the attacker requires less strategic complexity to counter a defender that only changes its actions once during match play, as opposed to a defender that is executing migration policies with structure on the scale of a few matches.

It is interesting to note the the attacker performs nearly as well against the \emph{SingleFlip-RandomOrder} defender, which randomly chooses between two available platform migration policies at the start of each generation, as it does against the more predictable \emph{SingleFlip-FixedOrder} defender, though the less predictable \emph{SingleFlip} defender causes a noisier attacker response. Examining the match level structure of the attacker response to the \emph{SingleFlip-RandomOrder} defender we find that the attacker has learned after a number of generations to observe the defender's move in \emph{match 1} of the game and predict all future OS activations by the defender, thus gaining the ability to efficiently counter the defender despite the random element in the defender's policy. This has implications to other active defense techniques, such as Address Space Layout Randomization (ASLR) \cite{shacham-2004}, for which randomization only increases the adversary's initial uncertainty and the entire system can be compromised given a single observation.

It is important to note that these results not only have implications to the attacker, who learns an optimal strategy against a specified defender, but also the defender as well. By evaluating the optimal fitness of an attacker against differing defenders, we also uncover the efficacy of different defensive strategies. If we assume the interaction is a zero-sum game -- any gains by the attacker are losses by the defender -- then the defender's goal should be to minimize the attacker fitness. 

\subsection{The 2-to-1 Game}

In the \emph{2-to-1} game the attacker faces a set of defenders that activate OS-A at a \emph{2-to-1} ratio to OS-B. As with the \emph{1-to-1} game, in the \emph{2-to-1} case the frequency of OS-A and OS-B activation is $2:1$ over the 365 matches taken as a whole, with the distribution of these activations among the matches differing among defender types. 

\begin{figure}[t]
\center
\includegraphics[width=9.0cm]{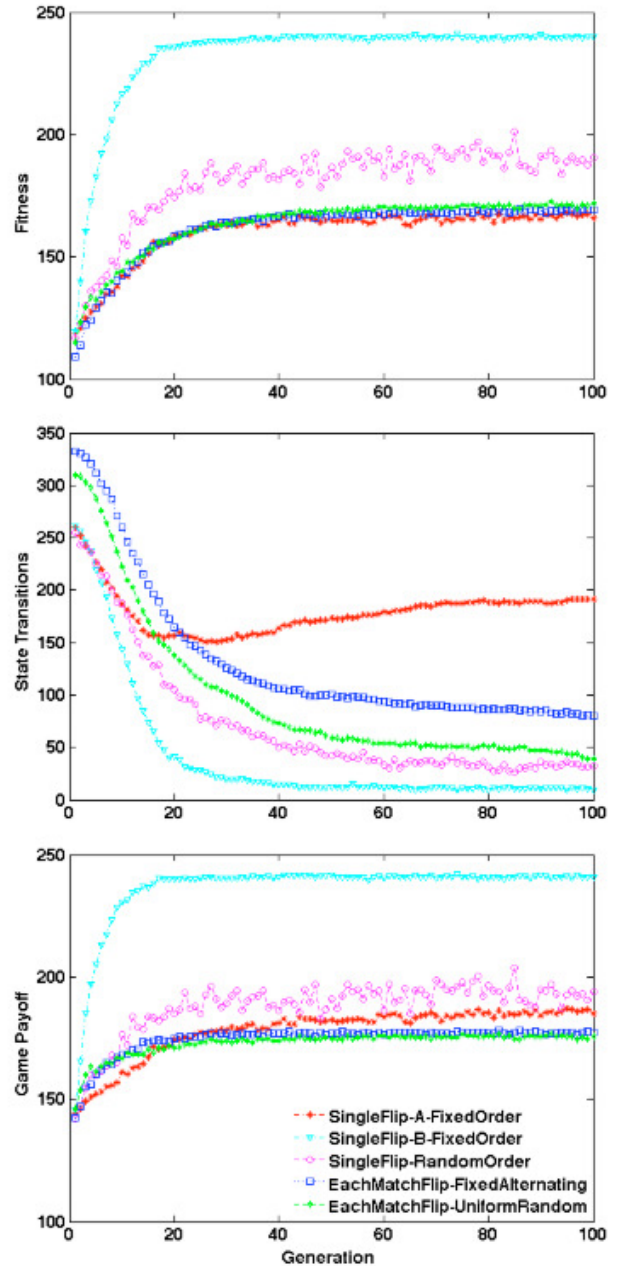}  
\caption{Attacker fitness (upper), state transitions (middle), and game payoff (lower) averaged over the 30 agent population and 100 simulation runs for the 5 defenders playing \emph{2-to-1} strategies.}
\label{2to1-triple}
\end{figure}

The evolved attacker responses to these defender policies are shown in Fig.\ref{2to1-triple}. Here the \emph{SingleFlip-A-FixedOrder} defender activates OS-A monolithically for the first 243 matches of a game, then switches to OS-B for the remaining 122 matches. On the other hand, the \emph{SingleFlip-B-FixedOrder} defender begins match play by activating OS-B for 122 consecutive matches, then switches to OS-A for the remaining 243 matches of the game.  In reaction to the different members of the \emph{2-to-1} family of defenders we see two different classes of responses by the attacker. The first, and simpler, type of response has the attacker deciding to essentially only invest in the creation of ZD-A, completely neglecting attacks against the less activated OS-B. The attacker uses this class of strategy against the \emph{EachMatchFlip-FixedAlternating} and \emph{EachMatchFlip-UniformRandom} defenders. We note in Fig. \ref{2to1-triple} that although \emph{a priori} this might seem a simple response by the attacker to biased defender policies, the strategic complexity cost is not negligible. We hypothesize that variability in the defender's OS activations on the scale of a few matches is an important factor in determining the transition structure (and so the strategic complexity cost) of the evolved attacker machines. This implies that a simple attacker response to a highly dynamic migration policy might entail a greater strategic complexity cost than would be expected from the attacker's simple actions alone. We will return to this question in a future study.

\begin{figure}[t]
\center
\includegraphics[width=9.0cm]{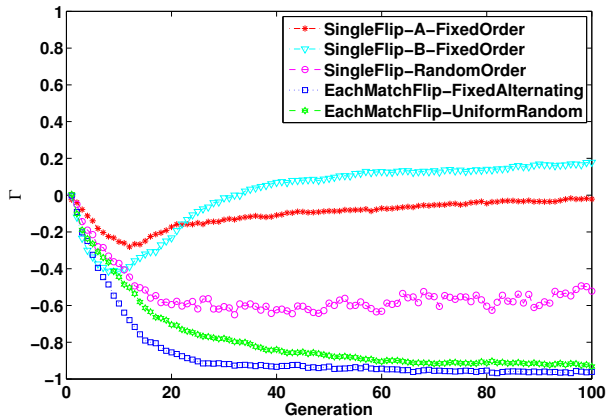} 
\caption{Attacker investment bias ($\Gamma$) for each of the platform migration policies in the \emph{2-to-1} defender family.}
\label{invBias-2to1}
\end{figure}

The second class of attacker response has the attacker making predictions about future defender activation activity and tuning resource investment accordingly. This prediction strategy is most successful against the \emph{SingleFlip-B-FixedOrder} defender, where the attacker learns to ignore the activation of OS-B in the first 122 matches of play, investing instead in ZD-A which has been brought into existence by the time the defender begins its 243 match run of OS-A activation. The tactical outcome of this attacker strategy bears resemblance to the simpler class of strategies discussed above in that only the ZD-A exploit ends up significantly benefiting the attacker, but the strategy underlying this result is quite different in the present case as evidenced by the distinctive investment bias result for the attacker facing the \emph{SingleFlip-B-FixedOrder} defender in Fig. \ref{invBias-2to1}. 

The attacker attempts a similarly intricate investment scheme when facing the \emph{SingleFlip-A-FixedOrder} defender, but to diminished success due primarily to the earlier noted inability to discern the true costs of exploit creation. When facing the \emph{SingleFlip-A-FixedOrder} defender the attacker learns to begin with focused investment in ZD-A to capitalize on the defenders early monolithic play of OS-A. The attacker continues investing heavily in ZD-A creation until the defender transitions to playing OS-B late in the game, investing well beyond the resources required to bring ZD-A into existence, resulting in a suboptimal result for the attacker. We speculate again that the learning algorithms underlying the attacker behavior are having difficulties grasping the fact that the attacker does not need to continue investing in an exploit after it has been created in order to benefit from its play against the defender. Again, we will take up this issue in a future study.

The \emph{SingleFlip-RandomOrder} defender randomly switches between the \emph{SingleFlip-A-FixedOrder} and the \emph{SingleFlip-B-FixedOrder} policies in each generation. The response to this more variable defender is similar to that in the \emph{1-to-1} case, in which the attacker strategy is wholly defined based on the observations of the initial activation by the defender. The defender's choice of which migration to activate in each round varies randomly between simulation runs, causing the aggregate result shown in Fig. \ref{2to1-triple} to be noisy (and reflective of the actual strategies being employed by the attacker only in a very approximate sense). 

\section{Conclusion}\label{S:Conclusion}

We have developed a model of adaptive strategy formulation and used it to study the choices an attacker learns to make to overcome temporal platform migration moving target defense strategies. The attacker-defender interaction has been modeled as a game in which a non-adaptive defender deploys a temporal platform migration defense. Against this active defense, a population of attackers developed strategies specifying the temporal ordering of resource investments that bring zero-day exploits into existence to compromise the defender's system.

We note that in this study, the attacker has often ignored the effects of causality by learning over generations which platforms the defender will play before they are actually played in each game. This actually maps to realistic scenarios, when an attacker is able to perform reconnaissance to learn a strategy before engaging a system. Random strategies do not suffer nearly as strongly as those which are predictable, suggesting that increased uncertainty, even with a defined strategy, improves defender performance.

This work has strong implications in cyber security, as it enables defenders to understand their defensive posture without having to explicitly detail an attacker. The reality is that defenders often do not know their adversaries' strategies, but they do know some of their goals, which can be codified into a fitness function. While we have demonstrated a simple case study in this paper, the framework we have developed is highly extensible and will be used to answer very detailed questions. In future work, we aim to expand this study by including complex defender strategies, numerous platforms, and the ability for adversaries to develop exploits that work against multiple systems.

\section*{Acknowledgments}

We thank William Streilein, Seth Webster, and Neal Wagner of MIT Lincoln Laboratory for helpful discussions.

\end{document}